\documentclass[conference]{IEEEtran}

\usepackage[
  pass,
]{geometry}

\usepackage[latin1]{inputenc}
\usepackage{amsmath,bbm,subfigure}

\usepackage{graphicx}
\usepackage{acronym}
\usepackage{changes}
\usepackage{lipsum}
\definechangesauthor[name={Authors}, color=blue]{authors}

\usepackage{acronym}

\usepackage[acronym]{glossaries}
\usepackage{glossary-mcols}
\usepackage{url,comment}
\usepackage{booktabs}
\usepackage{flushend}

\usepackage{amsmath}
\usepackage{algorithm}
\usepackage[noend]{algpseudocode}
\usepackage{siunitx} 
\usepackage{caption}

\usepackage{array}
\newcolumntype{L}[1]{>{\raggedright\let\newline\\\arraybackslash\hspace{0pt}}m{#1}}
\newcolumntype{C}[1]{>{\centering\let\newline\\\arraybackslash\hspace{0pt}}m{#1}}
\newcolumntype{R}[1]{>{\raggedleft\let\newline\\\arraybackslash\hspace{0pt}}m{#1}}

\usepackage{tikz}
\usepackage{pgfplots}
\usepackage{subfig}
\usetikzlibrary{shapes}
\usetikzlibrary{backgrounds}
\usetikzlibrary{calc}
\usetikzlibrary{positioning}
\usepgfplotslibrary{groupplots} 
\usetikzlibrary{pgfplots.groupplots}

\usepackage{multirow}
\usepackage{lmodern}
\makeatletter
\def\BState{\State\hskip-\ALG@thistlm}
\makeatother

\newacronym{AAA}{AAA}{Authentication, Authorization, and Accounting}
\newacronym{RSC}{RSC}{Recursive  Systematic Convolutional}
\newacronym{LLR}{LLR}{Log-Likelihood Ratio}
\newacronym{FS}{FS}{Functional Split}
\newacronym{BBU}{BBU}{Base Band Unit}
\newacronym{COTS}{COTS}{Commercial off-the-shelf }
\newacronym{VNF}{VNF}{Virtualized Network Function}
\newacronym{VNF FG}{VNF FG}{VNF Forwarding Graph}
\newacronym{NFV}{NFV}{Network Function Virtualization}
\newacronym{GPP}{GPP}{General Purpose Processor}
\newacronym{vEPC}{vEPC}{virtual Evolved Packet Core}
\newacronym{LTE}{LTE}{Long Term Evolution}
\newacronym{uRLLC}{uRLLC}{Ultra-Reliable Low-Latency Communications}
\newacronym{eMBB}{eMBB}{enhanced Mobile BroadBand}
\newacronym{mMTC}{mMTC}{massive Machine Type Communications}
\newacronym{OSM}{OSM}{Open Source MANO}
\newacronym{C-EPC}{C-EPC}{Cloud-EPC}
\newacronym{EPCaaS}{EPCaaS}{EPC as a Service}
\newacronym{TDD}{TDD}{Time Division Duplex}
\newacronym{UE}{UE}{User Equipment}
\newacronym{HARQ}{HARQ}{Hybrid Automatic Repeat-Request}
\newacronym{PRB}{PRB}{Physical Resource Blocks}
\newacronym{MCS}{MCS}{Modulation and Coding Scheme}
\newacronym{CQI}{CQI}{Channel Quality Indicator}
\newacronym{DC}{DC}{Dedicated Core}
\newacronym{RR}{RR}{Round Robin}
\newacronym{G}{G}{Greedy}
\newacronym{VPN}{VPN}{Virtual Private Network}
\newacronym{MPLS}{MPLS}{Multiprotocol Label Switching}
\newacronym{OWL}{OWL}{Web Ontology Language}
\newacronym{NST}{NST}{Network Slice Template}
\newacronym{NSST}{NSST}{Network Slice Subnet Template}
\newacronym{NSMF}{NSMF}{Network Slice Management Function}
\newacronym{NSSMF}{NSSMF}{Network Slice Subnet Management Function}
\newacronym{CSMF}{CSMF}{Communication Service Management Function}
\newacronym{FCAPS}{FCAPS}{Fault-management Configuration Accounting Performance and Security}
\newacronym{RF}{RF}{Radio Frequency}
\newacronym{RIC}{RIC}{RAN Intelligent Controller}
\newacronym{vBNG}{vBNG}{virtual Broadband Network Gateway}
\newacronym{BCR-NMS}{BCR-NMS}{Bootstrapper of Cloud-RAN - Network Management System}
\newacronym{CNF}{CNF}{Cloud-Native Network Function}
\newacronym{PLMN}{PLMN}{Public Land Mobile Network}
\newacronym{CLAMP}{CLAMP}{Closed Loop Automation Management Platform}
\newacronym{ODA}{ODA}{Open Digital Architecture}
\newacronym{FDD}{FDD}{Frequency Division Duplex}
\newacronym{OFDM}{OFDM}{Orthogonal Frequency Division Multiplexing}

\newacronym{VM}{VM}{Virtual Machine}
\newacronym{PDCP}{PDCP}{Packet Data Convergence Protocol}
\newacronym{MAC}{MAC}{Medium Access Control}
\newacronym{RLC}{RLC}{Radio Link Control}
\newacronym{RRC}{RRC}{Radio Resource Control}
\newacronym{AM}{AM}{Acknowledged Mode}
\newacronym{UM}{UM}{Unacknowledged Mode}
\newacronym{TM}{TM}{Transparent Mode}
\newacronym{MIMO}{MIMO}{Multiple Input Multiple Output}
\newacronym{MISO}{MISO}{Multiple Input Single Output}
\newacronym{SIMO}{SIMO}{Single Input Multiple Output}
\newacronym{SISO}{SISO}{Single Input Single Output}
\newacronym{MCC}{MCC}{Mobile Country Code}
\newacronym{MNC}{MNC}{Mobile Network Code}
\newacronym{S-TMSI}{S-TMSI}{Shortened Temporary Mobile Subscriber Identity}
\newacronym{IMSI}{IMSI}{International Mobile Subscriber Identity}
\newacronym{DRB}{DRB}{Dedicated Radio Bearer}
\newacronym{GUMMEI}{GUMMEI}{Globally Unique MME Identity}

\newacronym{PCI}{PCI}{Physical-layer Cell Identity}
\newacronym{ROHC}{ROHC}{Robust Header Compression}
\newacronym{SN}{SN}{Sequence Number}
\newacronym{RAR}{RAR}{Random Access Response}
\newacronym{C-RNTI}{C-RNTI}{Cell Radio Network Temporary Identifier}
\newacronym{BSR}{BSR}{Buffer Status Report}
\newacronym{DRX}{DRX}{Discontinuous Reception}
\newacronym{PHR}{PHR}{Power Head Room}
\newacronym{PUSCH}{PUSCH}{Physical Uplink Shared Channel}
\newacronym{ADM}{ADM}{Activation/Deactivation MAC}
\newacronym{GP}{GP}{Gap Period}

\newacronym{RE}{RE}{Resource Element}
\newacronym{RB}{RB}{Resource Block}
\newacronym{REG}{REG}{Resource Element Group}
\newacronym{CSRS}{CSRS}{Cell-Specific Reference Signal}
\newacronym{IFFT}{IFFT}{Inverse Fast Fourier Transform}
\newacronym{OFDMA}{OFDMA}{Orthogonal Frequency Division Multimple Access}
\newacronym{CRC}{CRC}{Cyclic Redundancy Check}
\newacronym{SFC}{SFC}{Service Function Chain}

\newacronym{eNB}{eNB}{Evolved NodeB}
\newacronym{RAN}{RAN}{Radio Access Network}
\newacronym{ARQ}{ARQ}{Automatic Repeat reQuest}
\newacronym{NAS}{NAS}{Non-Access Stratum}
\newacronym{MME}{MME}{Mobility Management Entity}
\newacronym{MIB}{MIB}{Master Information Block}
\newacronym{SIB}{SIB}{System Information Block}
\newacronym{RSRP}{RSRP}{Reference Signal Received Power}
\newacronym{RAT}{RAT}{Radio Access Technologie}
\newacronym{ACK}{ACK}{Acknowledge}
\newacronym{NACK}{NACK}{Negative acknowledge}
\newacronym{PDCCH}{PDCCH}{Physical Downlink Control Channel}
\newacronym{SAW}{SAW}{Stop and Wait}
\newacronym{TTI}{TTI}{Transmission Time Interval}
\newacronym{RRH}{RRH}{Radio Remote Head}
\newacronym{SNIR}{SNIR}{Signal-to-Noise-plus-Interference Ratio}
\newacronym{WCET}{WCET}{Worst Case Execution Time}
\newacronym{GPC}{GPC}{General Purpose Computer}
\newacronym{KPI}{KPI}{Key Performance Indicator}
\newacronym{OAI}{OAI}{Open Air Interface}
\newacronym{IMS}{IMS}{IP Multimedia Subsystem}
\newacronym{vIMS}{vIMS}{virtual IP Multimedia Subsystem}
\newacronym{EPC}{EPC}{Evolved Packet Core}
\newacronym{SDN}{SDN}{Software Defined Network}
\newacronym{C-RAN}{C-RAN}{Cloud-RAN}
\newacronym{OS}{OS}{Operating System}
\newacronym{TB}{TB}{Transport Block}
\newacronym{TBS}{TBS}{Transport Block Size}
\newacronym{QCI}{QCI}{QoS Channel Indicator}

\newacronym{GPU}{GPU}{Graphics Processing Unit}
\newacronym{CPU}{CPU}{Central Processing Unit}

\newacronym{SDU}{SDU}{Service Data Unit}
\newacronym{CBS}{CBS}{Code Block Size}
\newacronym{CB}{CB}{Code Block}
\newacronym{SPMD}{SPMD}{Single Program Multiple Data}
\newacronym{SIMD}{SIMD}{Single Instruction Multiple Data} 

\newacronym{SINR}{SINR}{Signal-to Interference Noise Ratio}
\newacronym{CO}{CO}{Central Office}
\newacronym{CA}{CA}{Carrier Aggregation}
\newacronym{SRS}{SRS}{Sound Reference Signal}
\newacronym{SC-OFDMA}{SC-OFDMA}{Single Carrier - Orthogonal Frequency Division Multiple Access}
\newacronym{FPGA}{FPGA}{Field-Programmable Gate Array}
\newacronym{TA}{TA}{Time Advancing}
\newacronym{CoMP}{CoMP}{Coordinated Multi-point}
\newacronym{NPRB}{NPRB}{Number of Physical Resource Blocks}
\newacronym{RTT}{RTT}{Round Trip Time}
\newacronym{CPRI}{CPRI}{Common Public Radio Interface}
\newacronym{CBR}{CBR}{Constant Bit Rate}
\newacronym{NRB}{NRB}{Number of Resource Blocks}
\newacronym{BJF}{BJF}{Biggest Job First}
\newacronym{EDF}{EDF}{Earliest Deadline First}
\newacronym{FCFS}{FCFS}{First-come, First-served}
\newacronym{PSTN}{PSTN}{Public Switched Telephone Network}
\newacronym{ETSI}{ETSI}{European Telecommunications Standards Institute}
\newacronym{vBBU}{vBBU}{virtualized BBU}
\newacronym{vRAN}{vRAN}{virtualized RAN}
\newacronym{IoT}{IoT}{Internet of Things}
\newacronym{B2B}{B2B}{Business to Business}
\newacronym{B2C}{B2C}{Business to Customer}
\newacronym{QoE}{QoE}{Quality of Experience}
\newacronym{QoS}{QoS}{Quality of Service}
\newacronym{VNO}{VNO}{Virtual mobile Network Operator}
\newacronym{SLA}{SLA}{Service Level Agreement}
\newacronym{VRRM}{VRRM}{Virtual Radio Resource Management}
\newacronym{KVM}{KVM}{Kernel-based Virtual Machine}
\newacronym{LXC}{LXC}{Linux Containers}
\newacronym{PS}{PS}{Processor Sharing}

\newacronym{eCPRI}{eCPRI}{evolved CPRI}
\newacronym{RoE}{RoE}{Radio over Ethernet}
\newacronym{PAPR}{PAPR}{Peak-to-average power ratio}
\newacronym{SC-FDMA}{SC-FDMA}{Single Carrier Frequency Division Multiple Access}
\newacronym{AGC}{AGC}{Automatic Gain Control}
\newacronym{PMD}{PMD}{Polarization Mode Dispersion}
\newacronym{ADC}{ADC}{Analogic-Digital Converter}

\newacronym{IQ}{IQ}{In-Phase Quadrature}
\newacronym{xRAN}{xRAN}{extensible Radio Access Network}
\newacronym{ISI}{ISI}{Inter-symbol interference}
\newacronym{FFT}{FFT}{Fast Fourier Transform}
\newacronym{IPC}{IPC}{Inter process communication}
\newacronym{CCDU}{CCDU}{Channel Coding Data Unit}
\newacronym{CC}{CC}{Channel Coding}
\newacronym{gNB}{gNB}{next-Generation Node B}
\newacronym{EUTRAN}{EUTRAN}{Evolved Universal Terrestrial Radio Access Network}
\newacronym{SCTP}{SCTP}{Stream Control Transmission Protocol}
\newacronym{NR}{NR}{New Radio}
\newacronym{NF}{NF}{Network Function}
\newacronym{CU}{CU}{Central Unit}
\newacronym{DU}{DU}{Distributed Unit}
\newacronym{NGC}{NGC}{Next Generation Core}
\newacronym{DL}{DL}{down-link}
\newacronym{UL}{UL}{up-link}
\newacronym{LJF}{LJF}{Largest Job First}
\newacronym{RANaaS}{RANaaS}{RAN as a Service}
\newacronym{NaaS}{NaaS}{Network as a Service}
\newacronym{NS}{NS}{Network Service}
\newacronym{FG}{FG}{Forwarding Graph}
\newacronym{VNFC}{VNFC}{VNF Component}
\newacronym{MANO}{MANO}{Management and Orchestration}
\newacronym{FIFO}{FIFO}{First In Firs Out}
\newacronym{NFVI}{NFVI}{NFV Infrastructure}
\newacronym{NFVO}{NFVO}{NFV Orchestrator}
\newacronym{PoP}{PoP}{Point of Presence}
\newacronym{NAT}{NAT}{Network Address Translation}
\newacronym{CDN}{CDN}{Content Delivery Network}
\newacronym{VNFM}{VNFM}{VNF Manager}
\newacronym{EM}{EM}{Element Management}
\newacronym{VIM}{VIM}{Virtualised Infrastructure Manager}

\newacronym{e2e}{e2e}{end-to-end}

\newacronym{AMF}{AMF}{Access and Mobility Management Function}
\newacronym{SMF}{SMF}{Session Management Function}
\newacronym{UPF}{UPF}{User Plane Function}
\newacronym{PCF}{PCF}{Policy Control Function}
\newacronym{UDM}{UDM}{Unified Data Management}
\newacronym{NRF}{NRF}{NF Repository Function}
\newacronym{AUSF}{AUSF}{Authentication Server Function}
\newacronym{API}{API}{Application Programming Interface}
\newacronym{HSS}{HSS}{Home Subscriber Server}
\newacronym{PCRF}{PCRF}{Policy and Charging Rules Function}
\newacronym{SOA}{SOA}{Software-Oriented Architecture}
\newacronym{AKA}{AKA}{Authentication and Key Agreement}
\newacronym{AF}{AF}{Application Function}
\newacronym{NEF}{NEF}{Network Exposure Function}
\newacronym{NSSF}{NSSF}{Network Slice Selection Function}
\newacronym{NSSP}{NSSP}{Network Slice Service Profile}
\newacronym{VES}{VES}{Virtual Event Streaming}
\newacronym{NSSAI}{NSSAI}{Network Slice Selection Assistance Information}

\newacronym{NSSI}{NSSI}{Network Slice Subnet Instance}

\newacronym{NSS}{NSS}{Network Slice Subnet}
\newacronym{NSC}{NSC}{Network Slice Customer}
\newacronym{NSP}{NSP}{Network Slice Provider}
\newacronym{CSC}{CSC}{Communication Service Customer}
\newacronym{CSP}{CSP}{Communication Service Provider}
\newacronym{SST}{SST}{Slice/Service Type}
\newacronym{SD}{SD}{Slice Differentiator}
\newacronym{USRP}{USRP}{UE Router Selection Policy}
\newacronym{S-NSSAI}{S-NSSAI}{Single Network Slice Selection Assistance Information}
\newacronym{ONISTT}{ONISTT}{Open Net-centric Interoperability Standards for Training and Testing}
\newacronym{KB}{KB}{Knowledge Base}
\newacronym{NSI}{NSI}{Network Slice Instance}

\newacronym{VF}{VF}{Virtual Function}
\newacronym{VFC}{VFC}{Virtual Function Component}
\newacronym{CR}{CR}{Complex Resource}
\newacronym{PNF}{PNF}{Physical Network Function}
\newacronym{CP}{CP}{Connection Point}
\newacronym{VL}{VL}{Virtual Link}
\newacronym{SDC}{SDC}{Service Design and Creation}
\newacronym{ONAP}{ONAP}{Open Network Automation Platform}
\newacronym{VID}{VID}{Virtual Infrastructure Deployment}
\newacronym{VSP}{VSP}{Vendor Software Product}

\newacronym{WEF}{WEF}{Wireless Edge Factory}

\newacronym{DP}{DP}{Data Plane}
\newacronym{ECOMP}{ECOMP}{Enhanced Control Orchestration Management and Policy}

\newacronym{AAI}{AAI}{Active and Available Inventory}
\newacronym{SDNC}{SDNC}{Software Defined Network Controller}
\newacronym{SO}{SO}{Service Orchestrator}
\newacronym{APPC}{APPC}{Application Controller}
\newacronym{DCAE}{DCAE}{Data Collection Analytics and Events}
\newacronym{OOF}{OOF}{ONAP Optimization Framework}
\newacronym{OSS}{OSS}{Operation Support System}
\newacronym{BSS}{BSS}{Business Support System}
\newacronym{SOCKS}{SOCKS}{Secured Over Credential-based Keberos}
\newacronym{VVP}{VVP}{VNF Validation Program}
\newacronym{PDP}{PDP}{Policy Decision Point}
\newacronym{PEP}{PEP}{Policy Enforcement Point}
\newacronym{PCC}{PCC}{Policy Creation Component}
\newacronym{RU}{RU}{Remote Unit}
\newacronym{ORAN}{ORAN}{Open-RAN}
\newacronym{CFS}{CFS}{Customer Facing Service}
\newacronym{RFS}{RFS}{Resource Facing Service}

\newacronym{VLM}{VLM}{Vendor License Model}
\newacronym{OCN}{OCN}{Open Core Network}
\newacronym{NBI}{NBI}{Northbound Interface}
\newacronym{I/Q}{I/Q}{in-phase / in-quadrature}
\newacronym{CUPS}{CUPS}{Control User Plane Separation}
\newacronym{BCR-S}{BCR-S}{Bootstrapper of Cloud-RAN - Server}
\newacronym{BCR}{BCR}{Bootstrapper of Cloud-RAN}
\newacronym{BCR-C}{BCR-C}{Bootstrapper of Cloud-RAN - Client}

\newacronym{ZTC}{ZTC}{Zero Touch Commissioning}

\newacronym{MNO}{MNO}{Mobile Network Operator}
\newacronym{NMS}{NMS}{Network Management System}

\newglossarystyle{modsuper}{%
  \glossarystyle{super}%
}

\begin{document}
\title{Towards 6G zero touch networks: \\  The case of automated Cloud-RAN deployments }

\author{ISBN:978-1-6654-3162-0 (IEEE) \\
\\
\IEEEauthorblockN{  Bini Angui, Romuald Corbel, Veronica Quintuna Rodriguez, Emile Stephan } 
\IEEEauthorblockA{Orange Labs, 2 Avenue Pierre Marzin,  22300 Lannion, France}  }

\maketitle  
\begin{abstract}
The arrival of 6G technologies shall massively increase the proliferation of on-demand and ephemeral networks. Creating and removing customized networks on the fly notably requires fully automation to guarantee commissioning-time acceleration. In this paper, we address the deployment automation of an end-to-end mobile network, with special focus on RAN units (referred to as Cloud-RAN). The Cloud-RAN automation is especially challenging due to the strong latency constraints expected in 6G as well as the required management of physical antennas. To automatically instantiate a Cloud-RAN chain, we introduce a Zero Touch Commissioning (ZTC) model which performs resource discovery while looking for both antennas and computing capacity as near as possible to the targeted coverage zone. We validate the ZTC model by a testbed which deploys, configures and starts the network service without human intervention while using Kubernetes-based infrastructures as well as open-source RAN and core elements implementing the mobile network functions.

\end{abstract}


\section{Introduction}

Automation, edge computing and artificial intelligence are key enablers of 6G technologies. 6G use-cases particularly address sustainable development, massive twinning, tele-presence, robots and local trust zones~\cite{uusitalo2021hexa,deli1hexax}. The requirement of local coverage for temporary usage will rapidly increase to achieve ephemeral and massive video transmissions during sportive/cultural events or even to launch medical emergency systems for specialized tele-assistance. The hosting infrastructure of 6G technologies should autonomously determine the best location to deploy 6G network functions, in order to fulfill the service requirements.

6G cloud-native infrastructures require to support automated deployment pipelines, not only to hold zero touch~\cite{ETSI_ZTM} paradigms (i.e., fully automated procedures without human intervention) but moreover to dynamically meet the performance requirements of 6G services in terms of bandwidth, latency and computing capacity. Cloud infrastructures shall natively able to append resources to support the workload or to move specific services to other locations to fulfill latency or computing requirements.

6G cloud-native network services can be conceived as a chain of microservices running on cloud infrastructures. One of the most challenging network services intended to be fully cloudified is the \gls{RAN} of mobile 6G networks, referred to as Cloud RAN. The complexity of Cloud-RAN~\cite{cloudRAN_mec} systems is on the tiny latency requirements involved in the physical layer, i.e., on the base band processing of radio signals. It is worth noting that the time budget to build and transmit RAN subframes is one and two milliseconds in the downlink and uplink, respectively (ultra low latency 6G usecases shall even require more reduced time slots). 

New RAN architectures plan to split up the RAN functions, and to deport part of the radio processing higher in the network in order to have a common intelligence to various radio sites. The 3GPP~\cite{3GPP38_801} particularly aims to split the RAN functions into three units \gls{CU}, \gls{DU} and \gls{RU}, this latter placed near to antennas. ~\cite{ORANWG4-CUS,veroIWCMCLaurent}.
When considering split Cloud-RAN architectures~\cite{3GPP38_801} the time budget needs to be shared between the runtime of RAN functions and the transmission time between the distant RAN units. Furthermore, the distance between the \gls{RU} and \gls{DU} (referred to as fronthaul size) needs to be as long as possible to enable centralization benefits (i.e., DUs coordination) but short enough to meet the latency requirements (less than 1 millisecond). Various Cloud-RAN deployment architectures have been studied to determine the most adapted placement of CU, DU, RU (distance between them) units in the Cloud infrastructure (involving Regional, Edge, and Far Edge zones). O-RAN has particularly defined six deployment scenarios~\cite{ORANWG6_cloudarch} and specified four of them~\cite{ORANWG6_cloud}. Furthermore, adapted resources discovery mechanisms need to be held during the automated 6G Cloud-RAN set up, to avoid exceeding the transmission time budget.

As various challenges need to be addressed to  meet both latency and computing constraints when automatically deploying 6G Cloud-RAN systems, we introduce in this paper, a specialized mechanism referred to as \gls{BCR} to support the zero touch Cloud-RAN management. It involves the resource discovery (fulfilling both computing and latency constraints to cover the targeted zone), synchronization and automated configuration of \glspl{RU} and \glspl{DU}-\glspl{CU} which are deployed on distant Kubernetes-based nodes. The automated deployment can be launched by means of an API coming from upper orchestration and management layers or by using a dedicated \gls{NMS}. 

As a proof of concept, we implement the proposed ZTC mechanism and the NMS including a graphic interface to create, delete and monitor deployments and resources. We validate the ZTC model while using open source Cloud-RAN units~\cite{veroIWCMCLaurent,veroNoF2020Romuald} developed on the basis of \gls{OAI}~\cite{oaiWebSite} code. We finally evaluate the performance of the proposed model in terms of agility and resource consumption in a fully cloudified infrastructure. Results demonstrate energy and resource consumption efficiency when running and deploying the proposed zero touch model. It is notably in line with 6G goals concerning sustainable development.

This paper is organized as follows: In Section~\ref{sec:arch}, we describe the architectural framework, as well as the placement patterns to be adopted when deploying Cloud RAN systems. The ZTC model and the various automated deployment stages are described in Section~\ref{sec:ZTC}. In Section~\ref{sec:proof}, we exhibit the testbed, while the performance evaluation is presented in Section~\ref{sec:perf}. Main conclusions are finally presented in Section~\ref{sec:conclusion}.

\section{Architecture Patterns for 6G Cloud-RAN at the Edge Network}
\label{sec:arch}
6G network services requiring low latency aim to be placed as close as possible to users. Edge computing paradigm focuses on this principle and enables both services and network functions to be placed near to users to reduce latency and bandwidth. Cloud-native network functions as those of Cloud-RAN aim to be deployed on edge architectures, particularly supported by a cloud infrastructure manager as Kubernetes(K8s)~\cite{ORANWG6_cloudarch}.

\begin{figure}[bt]
    \centering
    \includegraphics[width=\columnwidth]{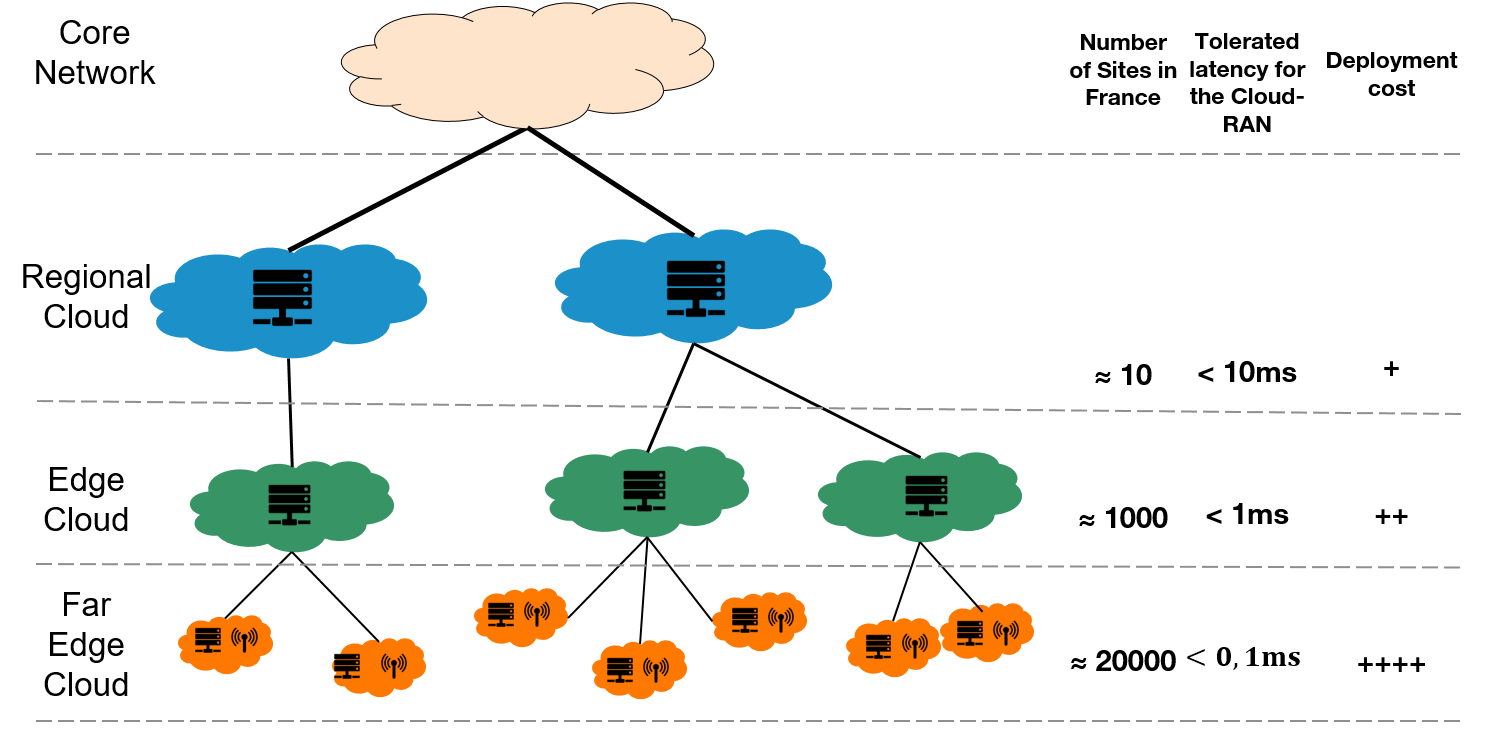}
    \caption{Substrate Network}
    \label{fig:archi-edge}
\end{figure}

The edge reference architecture for Cloud-RAN systems considers three levels of nodes, namely, \textit{Regional Cloud}, \textit{Edge Cloud} and \textit{Far Edge Cloud}. This topology (referred to as substrate network) has been notably adopted by O-RAN for the Cloud-RAN deployment~\cite{oranWebSite,ORANWG6_cloudarch}. Figure \ref{fig:archi-edge} illustrates these cloud areas.

Data centers (nodes) placed at the regional level allow centralizing network functions or application services that are not latency sensitive, e.g., \gls{vBNG}, \gls{UPF} or even the upper layer RAN functions. While regional nodes are placed around two hundred kilometers from end users ($10$ milliseconds), edge nodes are located only few tens of kilometers from them ($1$ millisecond). Some RAN functions as the radio scheduler and radio link controller are today clearly identified to be deployed at this level. Far edge nodes aim to host latency sensitive functions as those of the 6G RAN physical layer (e.g. modulation, demodulation).

 Various deployment scenarios intending to centralize or distribute RAN units have been widely studied in the literature, and considered by O-RAN~\cite{ORANWG6_cloudarch,ORANWG6_cloud}. In this work, we adopt the deployment of CU, DU, and RU in \textit{Regional}, \textit{Edge}, and \textit{Far Edge Cloud}, respectively, which is in line with O-RAN's \textit{Scenario F}. It is worth noting that the following \gls{ZTC} model is compliant with any deployment scenario.

 Deploying Cloud-RAN network functions at the Edge requires to address deployment challenges in the hosting infrastructure, i.e., K8s. Today, two main approaches can be used to deploy K8s at the edge: \textit{(i) }Distributed, which enables deploying the whole K8s cluster within edge nodes. Beyond the complexity of coordinating remote independent clusters, the drawback of distributed architectures is on the number of required replicas of control plane servers at each \textit{Edge Cloud} (e.g. 5 servers for a high availability cluster).  \textit{(ii)} Centralized, which keeps the control plane (master nodes) at the regional (central) cloud and manages remote edge nodes. The centralized deployment particularly allows to limit the number of control servers reducing CAPEX and OPEX. Thus, the centralized approach is in line with sustainable development goals expected in 6G. However, in the last years, separating worker nodes from their masters has been strongly avoided due to latency constraints between these two K8s elements. A recent study~\cite{Inria} demonstrates the feasibility of distancing worker nodes from masters.
In the following, we use the centralized approach as a basis to define the zero-touch deployment Cloud-RAN solution.

\section{ZTC Cloud-RAN Model}
\label{sec:ZTC}

\subsection{ZTC architectural framework }
\begin{figure}[bt]
    \centering

    \includegraphics[scale=0.5]{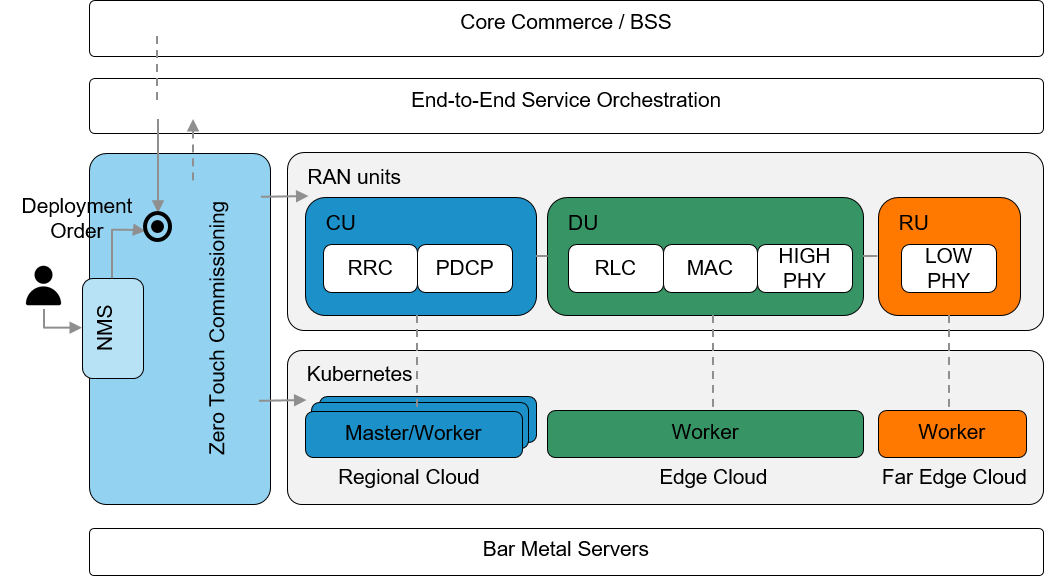}
    \caption{Architectural Framework}
    \label{fig:global}
\end{figure}

The general architectural framework of the proposed ZTC model is presented in Figure~\ref{fig:global}. It includes the upper layers (orchestration and core commerce), the hosting K8s cluster running on bar-metal servers and the RAN units implementing the various radio functions. 

The proposed ZTC procedure can be launched either from a \gls{MANO} platform, e.g., \gls{ONAP} or by using a dedicated \gls{NMS}. Thus, the orchestration service order requiring the instantiation of the various functions (containers) composing a Cloud-RAN system directly interacts with the K8s control elements, and notably with the proposed \gls{ZTC} module. The service order particularly contains the negotiated service level (e.g., geographic location, coverage area, number of supported users, data rate, etc) and the performance constraints to guarantee the correct functioning of the service, (e.g., minimum bandwidth between the hosting data-centers, maximum end-to-end latency, etc).

\subsection{Zero touch deployment procedure }

\begin{figure}[bt]
    \centering
    \includegraphics[width=\columnwidth]{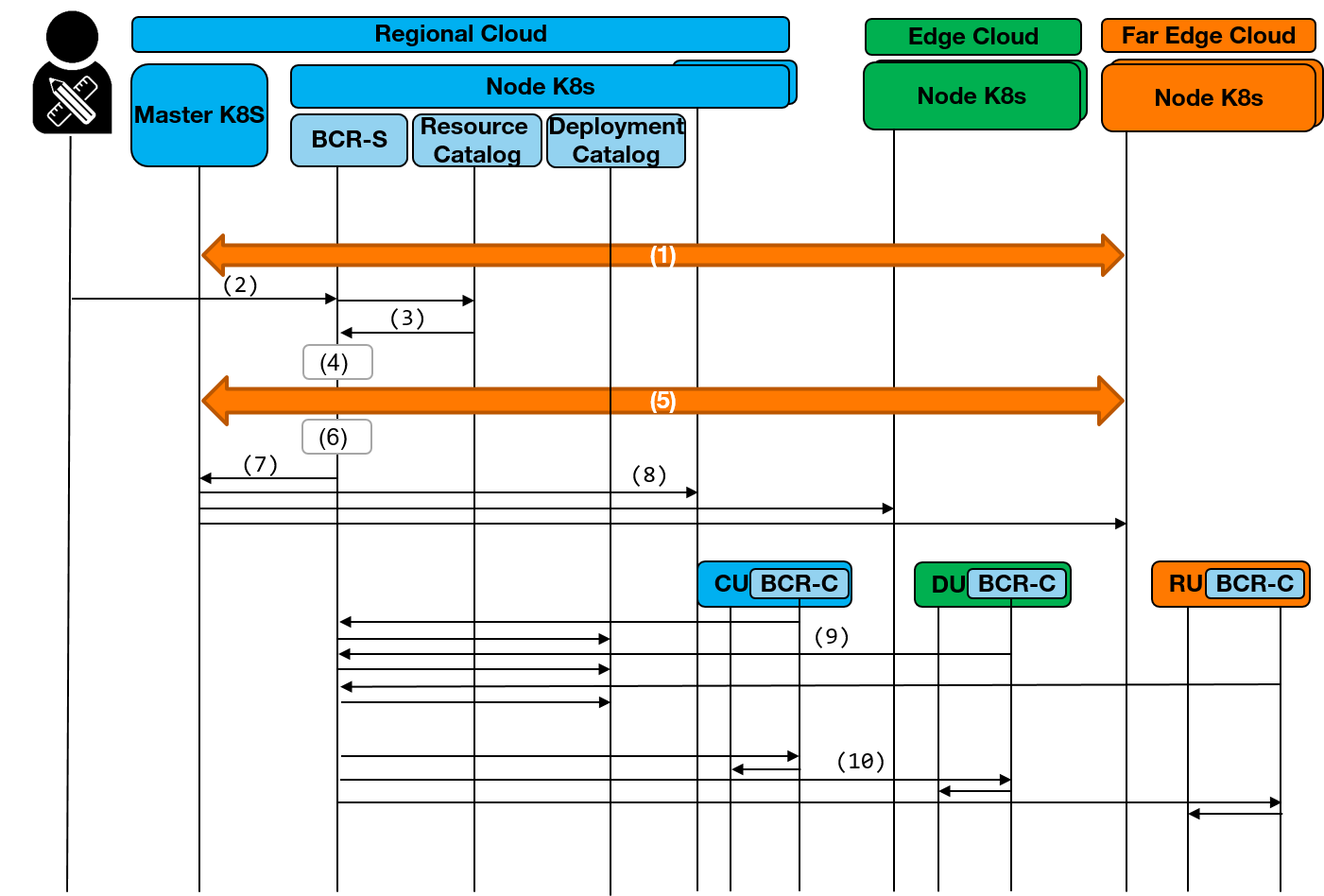}
    \caption{Call-flow of ZTC process}
    \label{fig:callFlow}
\end{figure}

To automate the Cloud-RAN deployment, we introduce in this work a specific control element referred to as \gls{BCR-S} which performs automated resource discovery and manages the ZTC procedures (instantiation, configuration, data base updates, etc). Independent control agents are also instantiated at each Cloud-RAN element (RU, DU, CU). These agents are referred to as \gls{BCR-C}. In addition to the control elements, the proposed automation system uses two catalogs, namely the Resource Catalog and the Deployment catalog, which respectively contain the information of K8s infrastructure (nodes, links) and the runtime information (e.g., IP addresses) of containers belonging to a given service (i.e., Cloud-RAN). The main steps involved in the automated Cloud-RAN deployment are (see Figure~\ref{fig:callFlow}):

    \paragraph*{ \textit{Building the Resource Catalog (1)}} The Resource catalog which is periodically updated by the BCR-S. It notably contains infrastructure nodes information (CPU, RAM, disk resources, geographical position, and connected antennas). The information stored in the catalog enables the validation of the Cloud-RAN requirements in terms of bandwidth, latency and computing capacity. This infrastructure catalog can be exposed to the orchestration layer (e.g. ONAP) in order to manage the lifecycle of end-to-end services.
     
    \paragraph*{ \textit{Service deployment order (2)}} The automated instantiation of Cloud-RAN elements starts with the service order coming from the orchestration layer or from the NMS. 
      
    \paragraph*{\textit{Resource discovery (3)}} The discovery function selects the potential hosting nodes that fulfill the requirements of the service order, i.e., geographic location, resource availability (antenna, RAM, CPU, disk), latency (e.g., the delay between RU and CU must be lower than 1 millisecond), and bandwidth (the required capacity of links connecting RUs and DUs which varies with the functional split and cell features).

    \paragraph*{\textit{List of best RU/DU/CU/antenna affiliations (4)}} While using the data obtained during the discovery stage, the BCR-S builds a list of the most adapted K8s nodes to host the RAN units RU, DU, CU according to the Cloud-RAN service order. It proposes one or more chains of K8s nodes that fullfil the affiliation requirements between the RAN units.
  
    \paragraph*{ \textit{Performance validation (5)}} The BCR-S provides a score to each proposed chain of nodes according to the fulfillment of the Cloud-RAN service requirements. The BCR-S launches the performance test for each chain of nodes while evaluating \gls{RTT} (end-to-end latency), hardware resources (nodes' capacity), and network throughput (available bandwidth in the optic fiber linking the nodes). The highest ranked chain of nodes will be then selected to host the service.  If no chain meets the requirements, the BCR-S aborts the deployment.

    \paragraph*{\textit{Automated Helm Charts creation (6)}} After performance validation, the BCR-S creates the required Helms charts (files used to define, install, and upgrade K8s applications) to deploy each of the RAN units, i.e., RU, DU, and CU, in the selected nodes. Helm charts uses the SLA parameters extracted from the service order (number of users, coverage, etc), as well as, the information of the hosting K8s nodes previously selected.

    \paragraph*{\textit{Launching the deployment of RAN units (7)}} While using created helm charts, the BCR-s triggers the deployment of RU, DU, CU units from the K8s Master node. 
   
    \paragraph*{\textit{Creation of RAN containers (8)}} The K8s master creates the containers of each RAN unit, RU, DU, CU according to the helm chart information. Each RAN container (RU, DU, CU) gets a dynamic IP address while using DHCP. A \gls{BCR-C} agent is deployed together with each RAN element to enable the ZTC. The BCR-C is then the entity that interacts with the BCR-S to affiliate (link) the RAN elements and to give rise to a complete gNodeB. 
    During the configuration of RUs, the BCR-S provides the serial number of the selected antenna to the RU's BCR-C. This latter sets up the antenna parameters in the RU configuration file.

    \paragraph*{ \textit{Getting and saving the information of RAN containers deployment (9)}} After the container creation, all information related to the deployment (e.g. IP addresses of RAN units and nodes) is sent to the deployment catalog. This latter stores the information of each RAN unit and enables the coordination with each other during their lifecycle (deployment, removal, monitoring, activation, deactivation, etc). 
    
    \paragraph*{\textit{Affiliating and starting RAN units (10)}} While using the information of the deployment catalog, the BCR-S provides to the BCR-C all the required parameters required to automatically configure and affiliate RAN units one to each other (RU needs to get the IP address of DU and this latter requires the RU IP, similarly, the DU and CU requires to know their respective IP addresses). Once configuration and affiliation is finished the BCR-C starts the \gls{RU}, \gls{DU}, and \gls{CU} applications.

\subsection{Testbed architecture}

\begin{figure}[bt]
    \centering
    \includegraphics[width=\columnwidth]{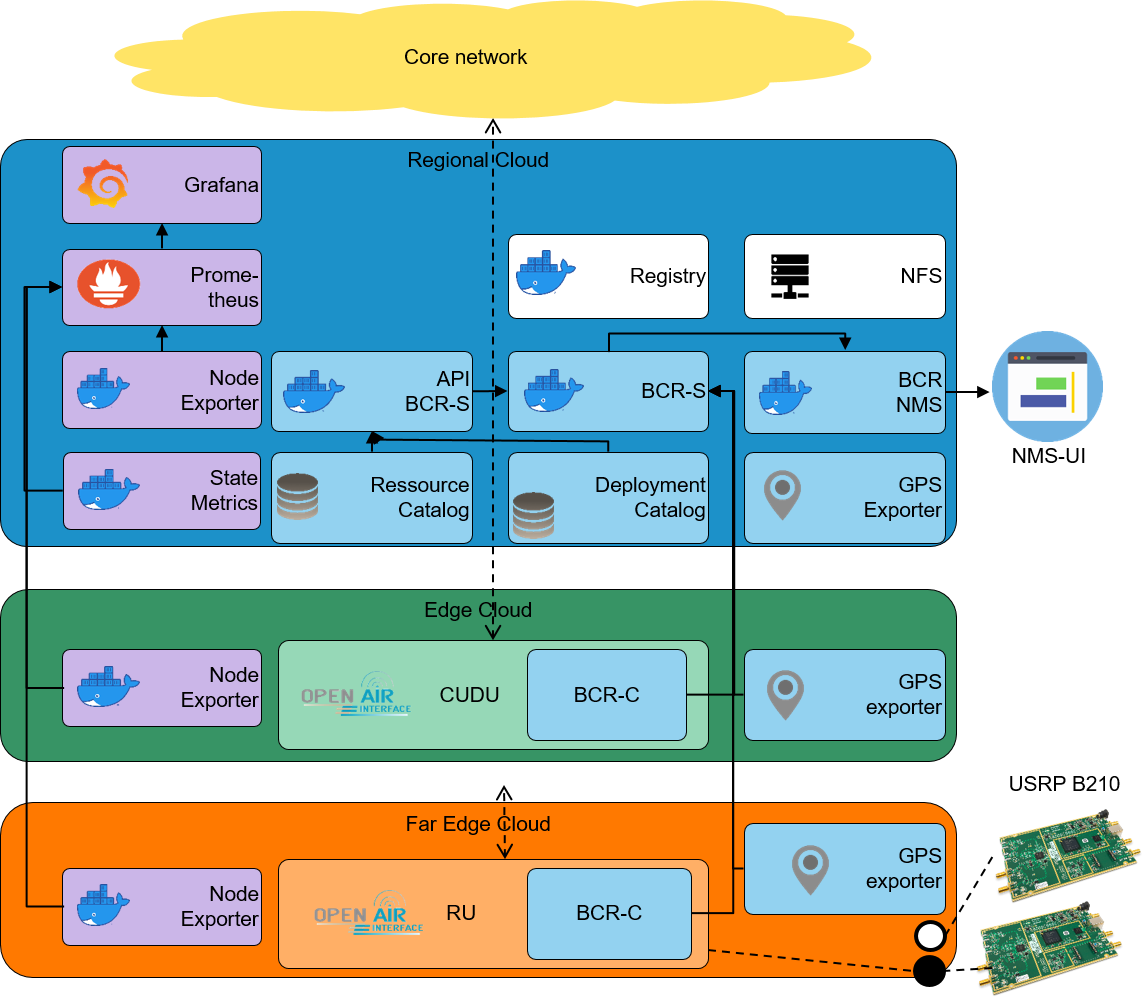}
    \caption{Testbed architecture}
    \label{fig:archi}
\end{figure}

The testbed architecture includes the K8s entities, the various control elements required for the Cloud-RAN zero touch provisioning and the RAN units spread into three cloud zones: \textit{Regional}, \textit{Edge} and \textit{Far Edge} respectively represented by three servers. Each cloud zone contains a single K8s worker node which is provisioned of a \textit{GPS exporter agent} in order to get the exact position of servers when automating the deployment as a function of coverage needs. Figure~\ref{fig:archi} illustrates the testbed architecture.

The \textit{Regional Cloud} hosts the main K8s entities, \textit{Ingress Controller} and \textit{Master node}, as well as the \textit{Private Registry} containing the RU, DU, CU images. RAN units implement the 7.3 functional split and have been developed by Orange on the basis of \gls{OAI}~\cite{veroNoF2020Romuald,veroIWCMCLaurent,icin2019} code.

The worker node of the \textit{Regional Cloud} hosts the ZTC control elements, namely \textit{Deployment Catalog}, \textit{Resource Catalog}, \textit{BCR-S}, as well as, an \textit{API BCR-S} (to interconnect one or more BCR-S to catalogs) and the \textit{BCR-NMS.} Central monitoring elements are also placed in the Regional worker node, namely, \textit{Prometheus}, \textit{Grafana}, \textit{State Metrics}, and \textit{Node exporter}. 

The \textit{Edge Cloud} hosts the CU-DU units and the corresponding BCR-C agent. The \textit{Far Edge Cloud} contains the RUs and their corresponding BCR-C agents, as well as two radio elements (USRP B210 cards) directly connected by USB port. 

Both \textit{Edge} and \textit{Far Edge} Cloud contain a \textit{node exporter} instance, which is used to collect metrics required for the monitoring. 

In order to evaluate the end-to-end service connection, the testbed integrates an OAI-generic core network. A user equipment can then be attached and traffic can be exchanged. 

\subsection{Functional experiments}

\subsubsection{Instantiating a Cloud-RAN chain}

When triggering the creation of a \textit{`New Cloud-RAN'} chain, various parameters are required in order to set up and to instantiate the RAN units, such as, the geographical area to be covered, the system name (referred to as \textit{tag}), the maximum number of supported users, etc. Configuration settings are either introduced by using the BCR-NMS, or sent into an API coming from orchestration and service management layers.

The BCR-S uses the settings to select the most adapted hardware (K8s nodes and antenna) to host the service chain (Cloud-RAN). Thus, the BCR-S creates the \textit{Helm charts} while using the identified nodes and received settings. Helm charts are triggered by the BCR-S from the K8s Master.

During the instantiation, the ZTC system, and more concretely the BCR-S and the BCR-C interact between each other in order to interconnect (setup network parameters) the RAN units. Thus, CU-DU registers the IP address of RU and vice-versa; antenna ports are also identified, When using OAI-based code, the selected antenna is specified by \emph{sdr\_addrs = "serial=SERIAL\_NUMBER"}.
 
The automation system (ZTC system) registers the deployment of the various containers (RAN units) in the Deployment Catalog and updates the system status while including servers/nodes occupancy, antennas availability in the Resource Catalog, Cloud RAN service health, coverage of the deployed RAN, etc.

\subsubsection{Cloud RAN Management}

The implemented ZTC system make available a NMS which notably includes a graphic interface to enable instantiating, deleting and managing Cloud-RAN systems.

\begin{figure}[bt]
    \centering
    \includegraphics[scale=0.22, trim=0 0 0 0, clip]{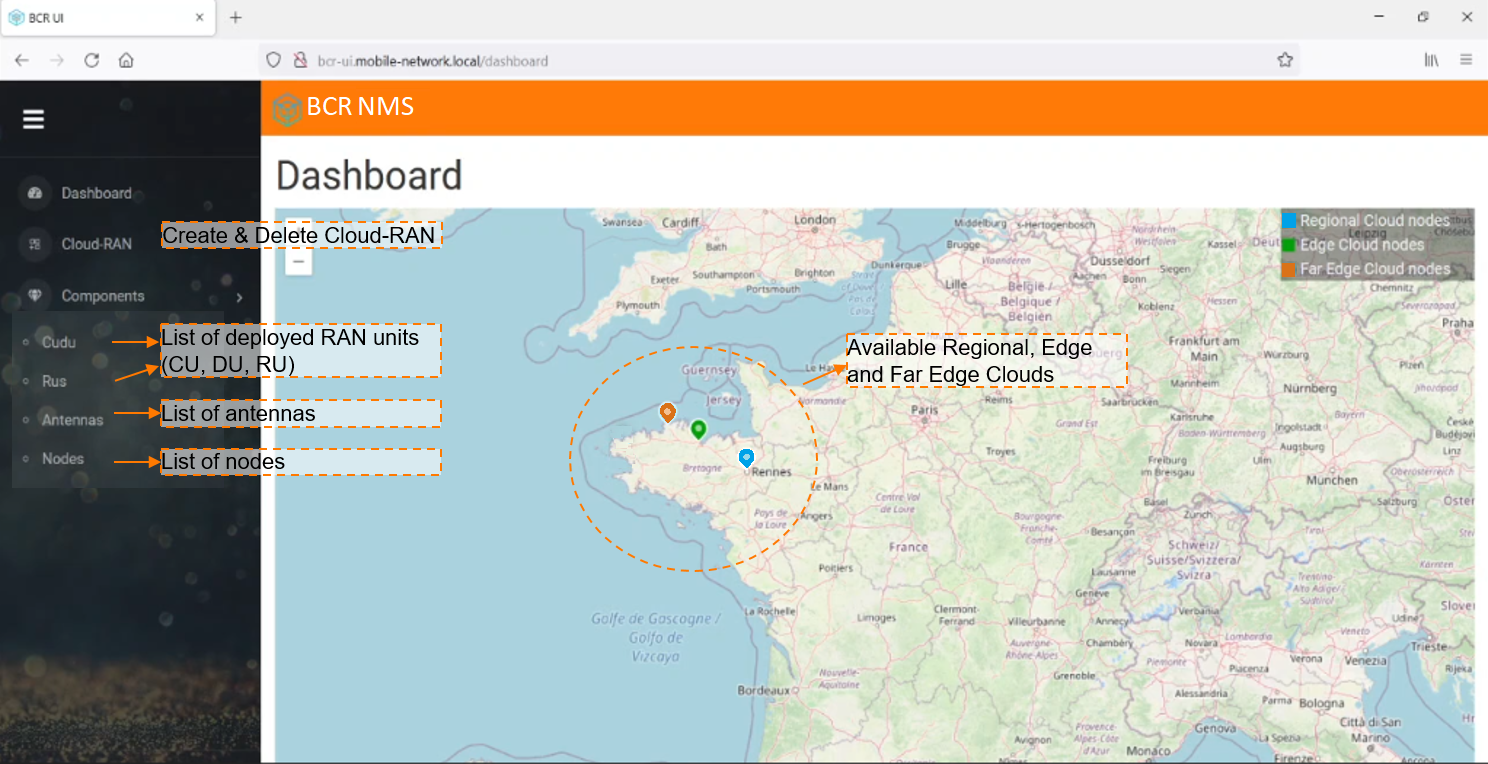}
    \caption{NMS - Available Clouds \& management options}
    \label{fig:ui-map}
\end{figure}

A map enables to locate in real-time all K8s nodes that are registered in the system, as well as, the deployed RAN units. The map additionally indicates the number of occupied/available antennas at each K8s node (See  Figure~\ref{fig:ui-map} for an illustration).
 
\section{Performance Evaluation}
\label{sec:perf}
We evaluate three main aspects: \textit{(i)} deployment time \textit{(ii)} hosting infrastructure health and \textit{(iii)} service (RAN units) health.  Main \glspl{KPI} are: deployment time, CPU and RAM consumption during deployment and execution.

\subsection{Deployment time}

While using Intel(R) Xeon(R) CPU E5-2640 v3 @ 2.60GHz, 125GB for Regional and Edge Cloud, and Intel(R) Core(TM) i7-8559U CPU @ 2.70GHz, 32 GB RAM for Far Edge Cloud, the experimented time to deploy the CU-DU and RU is $32$ seconds. This measured time includes the service start of each unit, the Cloud-RAN system is then ready to receive users connections. The deployment time of the ZTC module is $51$ seconds (only required once).

\subsection{Hosting infrastructure evaluation}
The (\textit{Regional}, \textit{Edge} and \textit{Far Edge Cloud}) evaluation is carried out by means of the deployed monitoring tools: \textit{Grafana}, \textit{Prometheus} and \textit{Node-Exporter}. These entities enable to obtain (in real-time) the resource consumption at each K8s node. 

We are particularly interested in evaluating the resource consumption before, during and after the deployment of both ZTC modules and Cloud RAN units. Thus, we define four main time indicators $t_1$, $t_2$, $t_3$ and $t_4$, which respectively identify the moment when the deployment of ZTC Helm charts is triggered from the K8s master node, the instant where the ZTC system starts running, the moment where the deployment of RAN units is launched and the instant when the creation of all RAN units is finished (i.e., Cloud-RAN execution starts). 
Figure~\ref{fig:courbe-CPU} illustrates the CPU consumption that is represented by the number of used CPU slots, while Figure~\ref{fig:courbe-RAM} shows the RAM occupancy in GigaBytes. Experimentation evidenced $t_1=12$, $t_2=63$, $t_3=138$ and $t_4=170$ seconds.

\def\courbeHorizontalSep{6em}
\def\courbeVerticalSep{5.5em}
\def\largeurCourbe{25em}
\def\hauteurCourbe{15em}
\def\colorCourbeUne{blue}
\def\colorCourbeDeux{green}
\def\colorCourbeTrois{orange}
\def\ymaxCPU{140}
\def\ymaxRAM{140}
\def\tun{12}
\def\tdeux{63}
\def\ttrois{138}
\def\tquatre{170}
\pgfplotsset{every axis title/.append style={at={(0.5,-0.5)}}}

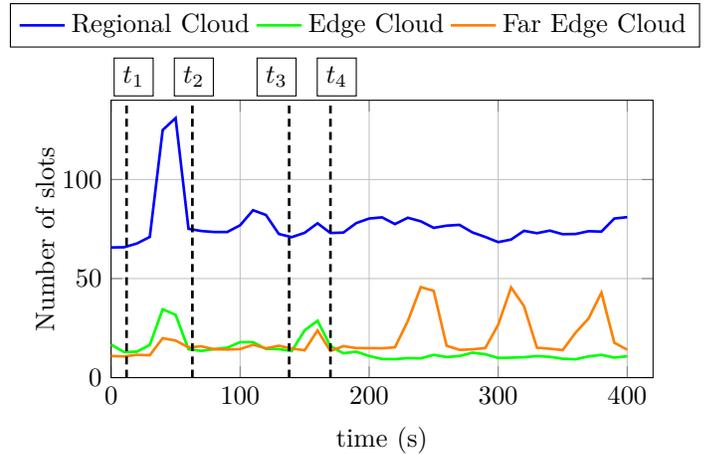
\begin{figure}
\centering
\begin{tikzpicture}

\node [draw] at (0.3,4) {$t_{1}$};
\node [draw] at (1.1,4) {$t_{2}$};
\node [draw] at (2.2,4) {$t_{3}$};
\node [draw] at (3,4) {$t_{4}$};

\centering
\begin{groupplot}[group style={group size=1 by 1, horizontal sep=\courbeHorizontalSep, vertical sep=\courbeVerticalSep}, width=\largeurCourbe, height=\hauteurCourbe]

\nextgroupplot[
	xlabel={time (s)},
	ylabel={Number of slots},
	xmin=0,
	xmax=420,
	ymin=0, 
	ymax=\ymaxCPU,
    ylabel style={yshift=-1em},
    legend style={at={(0.45,1.35)},
    anchor=north,
    legend columns=-1,},
    grid=both,
    grid style={line width=.1pt, draw=gray!10},
    major grid style={line width=.2pt,draw=gray!50},
]
\addplot[line width=1pt, color=\colorCourbeUne] table [x=Time, y=Regional-Cloud, col sep= semicolon, mark=none]{./courbes/cpu.csv};
\addplot[line width=1pt, color=\colorCourbeDeux] table [x=Time, y=Edge-Cloud, col sep= semicolon, mark=none]{./courbes/cpu.csv};
\addplot[line width=1pt, color=\colorCourbeTrois] table [x=Time, y=Far-Edge-Cloud, col sep= semicolon, mark=none]{./courbes/cpu.csv};
\addplot[line width=1pt, densely dashed, color=black] coordinates {(\tun,0) (\tun,\ymaxCPU)};
\addplot[line width=1pt, densely dashed, color=black] coordinates {(\tdeux,0) (\tdeux,\ymaxCPU)};
\addplot[line width=1pt, densely dashed, color=black] coordinates {(\ttrois,0) (\ttrois,\ymaxCPU)};
\addplot[line width=1pt, densely dashed, color=black] coordinates {(\tquatre,0) (\tquatre,\ymaxRAM)};
\addlegendentry{Regional Cloud}
\addlegendentry{Edge Cloud}
\addlegendentry{Far Edge Cloud}
\end{groupplot}
\end{tikzpicture}
    \caption{CPU consumption}
    \label{fig:courbe-CPU}
\end{figure}

\begin{figure}
\centering
\begin{tikzpicture}

\node [draw] at (0.3,4) {$t_{1}$};
\node [draw] at (1.2,4) {$t_{2}$};
\node [draw] at (2.2,4) {$t_{3}$};
\node [draw] at (3,4) {$t_{4}$};

\centering
\begin{groupplot}[group style={group size=1 by 1, horizontal sep=\courbeHorizontalSep, vertical sep=\courbeVerticalSep}, width=\largeurCourbe, height=\hauteurCourbe]
\nextgroupplot[
	xlabel={time (s)},
	ylabel={RAM (GB)},
	xmin=0,
	xmax=420,
	ymin=0,
	ymax=20,
    ylabel style={yshift=-1em},
    legend style={at={(0.45,1.35)},
    anchor=north,
    legend columns=-1,},
    grid=both,
    grid style={line width=.1pt, draw=gray!10},
    major grid style={line width=.2pt,draw=gray!50},
]
\addplot[line width=1pt, color=\colorCourbeUne] table [x=Time, y=Regional-Cloud, col sep= semicolon, mark=none]{./courbes/ram.csv};
\addplot[line width=1pt, color=\colorCourbeDeux] table [x=Time, y=Edge-Cloud, col sep= semicolon, mark=none]{./courbes/ram.csv};
\addplot[line width=1pt, color=\colorCourbeTrois] table [x=Time, y=Far-Edge-Cloud, col sep= semicolon, mark=none]{./courbes/ram.csv};
\addplot[line width=1pt, densely dashed, color=black] coordinates {(\tun,0) (\tun,\ymaxCPU)};
\addplot[line width=1pt, densely dashed, color=black] coordinates {(\tdeux,0) (\tdeux,\ymaxCPU)};
\addplot[line width=1pt, densely dashed, color=black] coordinates {(\ttrois,0) (\ttrois,\ymaxCPU)};
\addplot[line width=1pt, densely dashed, color=black] coordinates {(\tquatre,0) (\tquatre,\ymaxRAM)};
\addlegendentry{Regional Cloud}
\addlegendentry{Edge Cloud}
\addlegendentry{Far Edge Cloud}
\end{groupplot}
\end{tikzpicture}
    \caption{RAM consumption}
    \label{fig:courbe-RAM}
\end{figure}
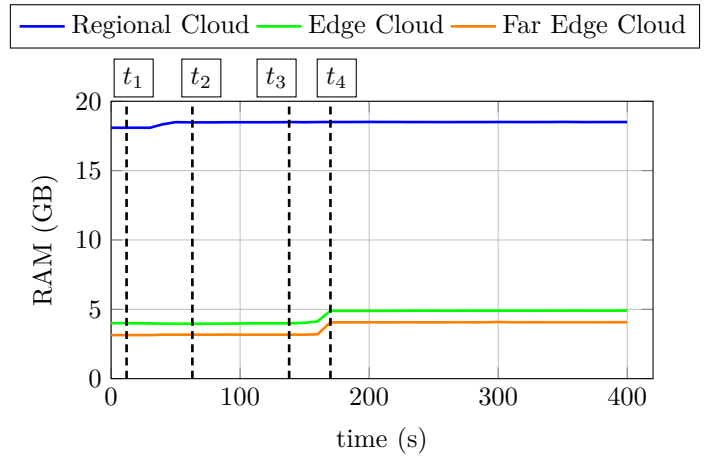

\paragraph*{ZTC system deployment ($t_1$-$t_2$)} 
 When the deployment in Regional Cloud is finished the CPU consumption drops back down to pre-deployment levels. Concerning \textit{Edge} and \textit{Far Edge} nodes, we observe a spike up in the CPU consumption due to the deployment of GPS simulators.

\paragraph*{ZTC operation ($t_2$-$t_3$)} 
The resource consumption during the runtime (operation) of the proposed automation and management system is negligible. The only significant variation is the CPU consumption observed in the \textit{Regional Cloud} during the update of databases containing K8s nodes, antennas, and deployments.

\paragraph*{RAN units deployment ($t_3$-$t_4$)} 

We can observe an increment of the CPU consumption at the three Cloud levels. In the \textit{Regional Cloud}, it is due to the BCR processing required to affiliate the RAN containers to the K8s hosting nodes. In \textit{Edge} and \textit{Far Edge} the CPU consumption increment is due to the deployment of CU-DU and RU, respectively.

\paragraph*{Cloud RAN operation (after $t_4$)} 

After the deployment of a new Cloud-RAN, the CPU usage in the \textit{Regional Cloud} slightly increases due to the operation of K8s.  In the \textit{Far Edge} node, the CPU consumption increases due to radio signal processing and antenna control (USRP B210). No significant increase in RAM is evidenced.

\subsection{End-to-end service performance evaluation}

In order to validate the end-to-end service availability, beyond the Cloud-RAN deployment, we perform the attachment and Connection of \gls{COTS} user equipments. We additionally validate the implemented multi-antenna management mechanism.

\begin{figure}[bt]
    \centering
    \includegraphics[scale=0.22, trim=0 50 0 0, clip]{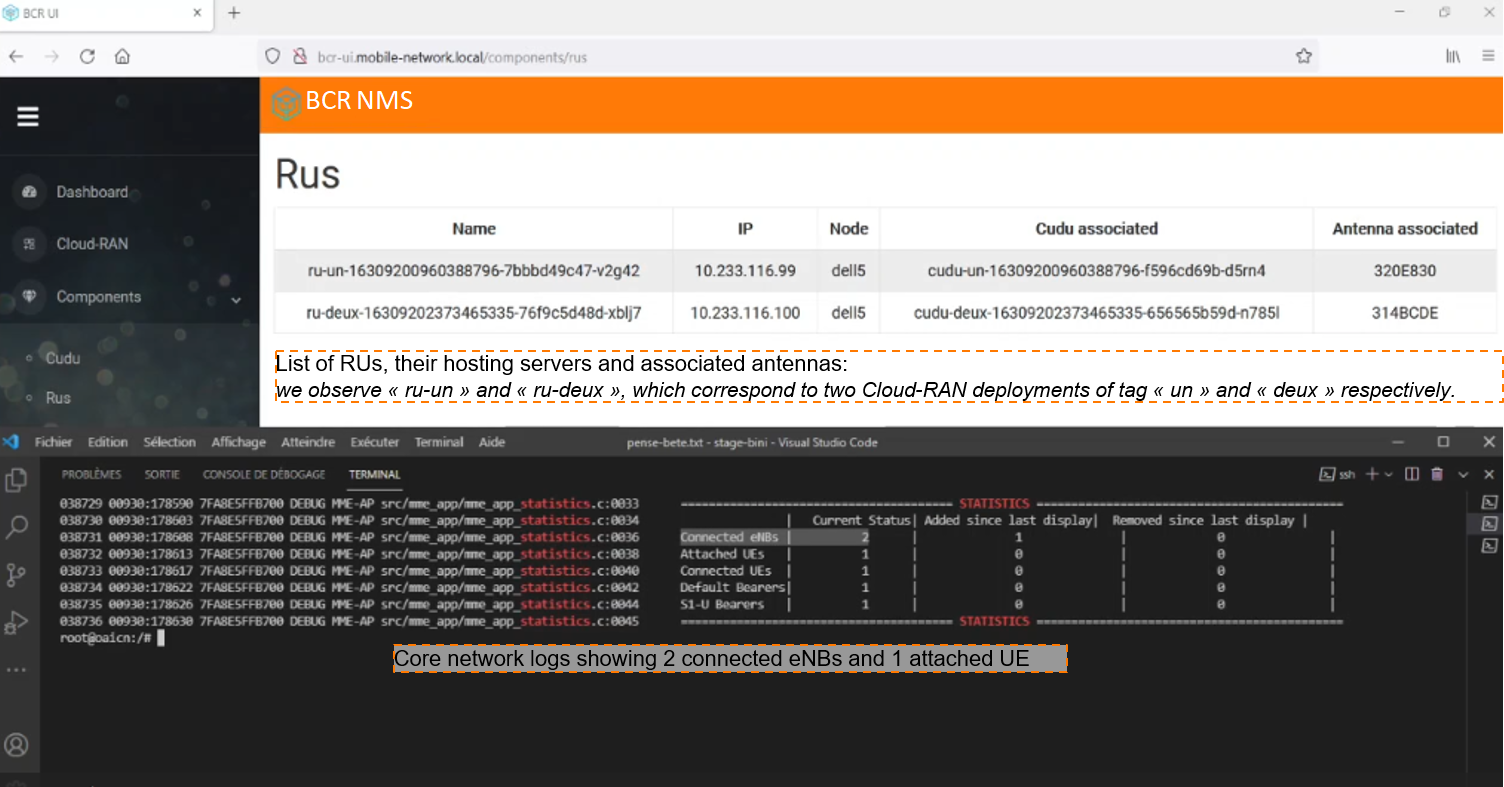}
    \caption{Core logs \& list of deployed RUs}
    \label{fig:logs}
\end{figure}

Figure~\ref{fig:logs} shows the list of the deployed RUs and the associated antennas, as well as, the core network logs showing two connected eNBs and a connected user to the automatically deployed mobile network.

\section{Conclusion}
\label{sec:conclusion}
We have addressed in this work one of the key enablers of incoming 6G services, i.e., the zero touch deployment of end-to-end services, particularly required for on-demand and ephemeral networks. As driving usecase we study the deployment automation of Cloud RAN systems, which are composed of CU, DU, RU units respectively placed in three cloud levels: \textit{Regional}, \textit{Edge} and \textit{Far Edge}. This deployment architecture is compliant with the `Scenario F' of O-RAN. We have notably proposed a Zero Touch Commissioning (ZTC) model which enables automated resource discovery (hosting servers and antennas) as well as automated lifecycle management of RAN units, including instantiation, configuration, monitoring and deletion. The implemented system makes available a dedicated Network Management System (NMS) with a graphical interface to facilitate the operation and management of Cloud-RAN systems. The Cloud-RAN deployment is then launched by a \textit{deployment order} coming either from the NMS or from the service orchestration layers. The deployment order includes main required Cloud-RAN settings such as the geographic area to be covered, the maximal number of users, spectrum band, among others service features. The ZTC system selects the adapted K8s nodes to host the RAN units (CU, DU, RU) and automatically configures them, i.e., it provides IP addresses to each unit, selects available antennas, opens the required ports to enable traffic exchange and starts the service.  We have additionally evaluated the case of multi-antenna Far Edge nodes in order to support TowerCos scenarios where various RUs can be instantiated in a single shared infrastructure (K8s node) placed near to towers.
Performance results evidence that the resource consumption of the proposed ZTC systems is negligible.  The automated deployment of a fully cloudified end-to-end RAN system only requires $32$ seconds without any human intervention, instead of various hours required for standard deployments. Users attachments and connections validated the end-to-end service availability. 

\bibliographystyle{IEEEtran}
\bibliography{biblo}

\end{document}